\documentclass[aps,twocolumn,pra]{revtex4}
\usepackage{amsmath,amssymb,bm}
\usepackage{graphicx}
\usepackage{epstopdf}
\usepackage{latexsym}
\usepackage{subfigure}
\usepackage[usenames,dvipsnames]{color}
\usepackage{hyperref}
\usepackage{natbib}

\newcommand {\be}{\begin{equation}}
\newcommand {\ee}{\end{equation}}
\newcommand {\ba}{\begin{eqnarray}}
\newcommand {\ea}{\end{eqnarray}}

\newcommand{\LiCs}{{$^{6}$Li$^{133}$Cs}}
\newcommand{\NaK}{$^{23}$Na$^{40}$K}
\newcommand{\RbCs}{$^{87}$Rb$^{133}$Cs}
\newcommand{\KRb}{$^{40}$K$^{87}$Rb}
\newcommand{\LiNa}{{$^{6}$Li$^{23}$Na}}

\begin{document}

\title{The Molecular Hubbard Hamiltonian: Field Regimes and Molecular Species}
\author{M. L. Wall$^1$\footnote{Present address: JILA, University of Colorado, Boulder, Colorado 80309-0440, USA}, Erman Bekaroglu$^1$, and Lincoln D. Carr$^{1,2}$}
\address{$^1$Department of Physics, Colorado School of Mines, Golden, CO 80401, USA}
\address{$^2$Universit\"at Heidelberg, Physikalisches Institut, D-69120 Heidelberg, Germany}

\begin{abstract}
The molecular Hubbard Hamiltonian (MHH) naturally arises for ultracold ground state polar alkali dimer molecules in optical lattices.   We show that, unlike ultracold atoms, different molecules display different many-body phases due to intrinsic variances in molecular structure even when the molecular symmetry is the same.  We also demonstrate a wide variety of experimental controls on $^1\Sigma$ molecules via external fields, including applied static electric and magnetic fields, an AC microwave field, and the polarization and strength of optical lattice beams.  We provide explicit numerical calculations of the parameters of the MHH, including tunneling and direct and exchange dipole-dipole interaction energies, for the molecules {$^{6}$Li$^{133}$Cs}, $^{23}$Na$^{40}$K, $^{87}$Rb$^{133}$Cs, $^{40}$K$^{87}$Rb, and {$^{6}$Li$^{23}$Na} in weak and strong applied electric fields.  As case studies of many-body physics, we use infinite-size matrix product state methods to explore the quantum phase transitions from the superfluid phase to half-filled and third-filled crystalline phases in one dimension.
\end{abstract}

\maketitle

\section{Introduction}
Molecules cooled to ultralow temperatures provide access to a wealth of exciting new physics, such as the study of chemical reactions at ultracold temperatures, precision tests of fundamental symmetries, novel means of quantum information processing, and many-body physics with many internal degrees of freedom~\cite{reviews}.  The vast tunability proffered by molecules has led to many proposals to manipulate molecules with external fields in order to simulate outstanding unsolved quantum Hamiltonians motivated by solid state physics~\cite{Michelietal,Gorshkov_Manmana,Manmana_Stoudenmire_12}.  Taking a complementary approach, we address the immediately accessible many-body physics observable in upcoming experiments using ultracold ground state $^1\Sigma$ molecules in optical lattices, in which external fields will not be fine-tuned.   In this work, we use the molecular Hubbard Hamiltonian (MHH) to show that the first experiments with ultracold molecules in optical lattices will display a variety of many-body phases which are strongly dependent on molecular species and amenable to control with external fields.

For ultracold alkali atoms, interactions are well-modeled by a contact pseudopotential proportional to the $s$-wave scattering length, and the $s$-wave scattering length is readily tunable by using a Feshbach resonance.  Hence, while the specific positions of Feshbach resonances may differ among different alkali species, essentially the same tunability of interactions is provided for all species.  We demonstrate that the same is not true for ultracold molecules.  Due to intrinsic variations in the structure of $^1\Sigma$ alkali dimer molecules, different molecules access different regimes of the MHH, and so display a variety of many-body phases.  In this work,  we take as examples the bosonic molecule \RbCs~\cite{RbCs} and the fermionic molecules \LiCs~\cite{LiCs}, \NaK~\cite{NaK}, \KRb~\cite{KRb}, and \LiNa~\cite{LiNa}, all of which are being explored in current experiments.  After identifying the MHH for a particular near-term experimental setup we discuss some of its experimental controls which are not available to ultracold atoms. We then provide explicit numerical values for the parameters appearing in a quasi one-dimensional reduction of the MHH such as tunneling and dipolar interaction terms.  To provide case studies of many-body physics, we numerically calculate quantum phase transitions from superfluid to crystalline phases for several different molecular species at rational densities in one dimension using variational matrix product state techniques.  By computing phase diagrams in terms of experimental parameters such as the strength of an applied electric field, we show that crystalline phases will not be seen for the molecules with the weakest dipole moments.

\section{The molecular Hubbard Hamiltonian} 
The molecular Hubbard Hamiltonian (MHH) in the simplest experimental configuration is
\begin{align}
\nonumber  \hat{H}&=-\sum_{i\ne j\sigma;r_t}t_{j-i,\sigma}\hat{a}_{i,\sigma}^{\dagger}\hat{a}_{j,\sigma}-\sum_{\sigma\sigma'}\Omega_{\sigma\sigma'}\sum_i\hat{a}_{i,\sigma}^{\dagger}\hat{a}_{i,\sigma'}\\
\label{eq:MHH} & +\sum_{i\sigma}\Delta_{\sigma}\hat{n}_{i,\sigma}+\frac{1}{2}\sum_{i\ne j\sigma\sigma';r_U}U_{j-i,\sigma\sigma'}\hat{n}_{i,\sigma}\hat{n}_{j,\sigma'}\\
 \nonumber &+ \frac{1}{4}\sum_{i\ne j\sigma_1\sigma_2\sigma_2'\sigma_1';r_E}E_{j-i,\sigma_1\sigma_2\sigma_2'\sigma_1'}\hat{a}^{\dagger}_{i,\sigma_1}\hat{a}^{\dagger}_{j,\sigma_2}\hat{a}_{j,\sigma_2'}\hat{a}_{i,\sigma_1'}\, .
\end{align}
Here $\sigma$ labels the internal state of a molecule, and the operator $\hat{a}_{i,\sigma}$ ($\hat{a}_{i\sigma}^{\dagger}$) represents either a hard-core bosonic or fermionic annihilation (creation) operator of a molecule in state $\sigma$ on lattice site $i$.  In order, the terms in Eq.~\eqref{eq:MHH} are tunneling of molecules between sites $i$ and $j$ with state-dependent tunneling energy $t_{j-i,\sigma}$, local transitions between molecular internal states $\sigma\to\sigma'$ driven by an AC microwave field $\mathbf{E}_{\mathrm{AC}}$ which has right-circular polarization in the lab frame and effective Rabi frequency $\Omega_{\sigma\sigma'}$, energetic detuning $\Delta_{\sigma}$ of a molecule in state $\sigma$ from a reference ground state, \emph{direct} dipole-dipole interactions $U_{j-i,\sigma\sigma'}$ between a molecule in state $\sigma$ on site $i$ and a molecule in state $\sigma'$ on site $j$, and \emph{exchange} dipole-dipole interactions $E_{j-i,\sigma_1\sigma_2\sigma_2'\sigma_1'}$ which exchange a quantum of rotation between molecules at sites $i$ and $j$ through the dipole-allowed transitions $\sigma_1'\to \sigma_1$ and $\sigma_2'\to\sigma_2$.    The parameters $r_{t}$, $r_{U}$, and $r_{E}$ represent cutoffs of the range of tunneling and direct and exchange dipole-dipole interactions, respectively, and set a consistent energetic order of approximation for the terms in Eq.~\eqref{eq:MHH} for a given set of field parameters.  We note that these cutoffs generally depend on the internal state, $r_t=r_{t,\sigma}$ etc., and on directionality, e.g.~the range of tunneling will be shorter along directions in which the lattice potential is deeper.

The Hubbard parameters in Eq.~\eqref{eq:MHH} are defined in terms of integrals involving the lowest band Wannier functions $w_{i\sigma}(\mathbf{r})$, where $\sigma$ is an internal state index and $i$ denotes the centering site of the Wannier function.  The Wannier functions $w_{i\sigma}(\mathbf{r})$ are the quasimomentum Fourier transforms of the lowest band Bloch functions
\begin{align}
\label{eq:Hsp}\hat{H}_{\mathrm{sp}}\psi_{\sigma\mathbf{q}}(\mathbf{r})&=E_{\sigma\mathbf{q}}\psi_{\sigma\mathbf{q}}(\mathbf{r})\, ,
\end{align}
which are the simultaneous eigenfunctions of the lattice translation operators and the single-particle Hamiltonian $\hat{H}_{\mathrm{sp}}$.  In particular, we define
\begin{align}
\label{eq:tdef}&t_{j-i,\sigma}\equiv -\int_{\mathrm{BZ}} {d\mathbf{q}} e^{-i\mathbf{q}\cdot(\mathbf{R}_i-\mathbf{R}_j)}E_{\sigma\mathbf{q}}/{v_{\mathrm{BZ}}}\, ,\\
\label{eq:Deltadef}&\Delta_{\sigma}\equiv -t_{0,\sigma}-\omega\delta_{\bar{N},1}\, ,\\
\label{eq:U}&U_{j-i,\sigma\sigma'} \equiv (\mathcal{U}^{0;ijji}_{\sigma\sigma'\sigma'\sigma}+\mathcal{S}^{ijji}_{\sigma\sigma'})\pm (\mathcal{U}^{0;ijij}_{\sigma\sigma'\sigma'\sigma}+\mathcal{S}^{ijij}_{\sigma\sigma'})\, ,\\
\label {eq:E} &E_{j-i,\sigma_1\sigma_2\sigma_2'\sigma_1'}\equiv \sum_{q=-1,1}\,[\mathcal{U}^{q;ijji}_{\sigma_1\sigma_2\sigma_2'\sigma_1'}\pm \mathcal{U}^{q;ijij}_{\sigma_1\sigma_2\sigma_2'\sigma_1'}]\, ,\\
\label{eq:Omegadef}&\Omega_{\sigma\sigma'}\equiv E_{\mathrm{AC}}\left(d_{1,\sigma\sigma'}+d_{1,\sigma'\sigma}\right)\,,\\
\label{eq:Omega} &d_{q,\sigma\sigma'}\equiv \int d\mathbf{r} w_{0\sigma}^{\star}(\mathbf{r}) \hat{d}_qw_{0\sigma'}(\mathbf{r})\, .
\end{align}
Here, $v_{\mathrm{BZ}}$ is the volume of the first Brillouin zone (BZ), $\mathbf{R}_i$ is the coordinate of lattice site $i$, $\hat{d}_q$ is the projection of the dipole operator along space-fixed spherical basis element $\mathbf{e}_q$, and the positive (negative) sign in Eqs.~\eqref{eq:U} and \eqref{eq:E} refers to bosons (fermions).  The presence of $\delta_{\bar{N},1}$ in Eq.~\ref{eq:Deltadef} denotes that we work in the frame in which all states in the first excited rotational manifold rotate with the frequency $\omega$ of the AC field~\cite{HMHH}.  Working in the rotating frame makes the coupling to the AC field, Eq.~\eqref{eq:Omegadef}, time-independent.  The dipole and short-range interaction integrals appearing in \eqref{eq:U} and \eqref{eq:E} are
\begin{align}
\label{eq:geo} \mathcal{U}^{q; i_1i_2i_2'i_1'}_{\sigma_1\sigma_2\sigma_2'\sigma_1'}=&\textstyle-2\int d\mathbf{r} d\mathbf{r}'f^{i_1i_1'}_{\sigma_1\sigma_1'}(\mathbf{r})\hat{d}_q\hat{d}_{-q} \frac{C_0^{(2)}(\mathbf{r}-\mathbf{r}')}{|\mathbf{r}-\mathbf{r}'|^3} f_{\sigma_2\sigma_2'}^{i_2i_2'}(\mathbf{r}')\, ,\\
\label{eq:Sgeo} \mathcal{S}^{i_1i_2i_2'i_1'}_{\sigma\sigma'}=& \int d\mathbf{r} d\mathbf{r}'f^{i_1i_2'}_{\sigma\sigma}(\mathbf{r})P\left(\bar{a}\right) f_{\sigma'\sigma'}^{i_2i_2'}(\mathbf{r}')\, ,
\end{align}
where $f_{\sigma\sigma'}^{ii'}(\mathbf{r})\equiv w_{i\sigma}^{\star}(\mathbf{r})w_{i'\sigma'}(\mathbf{r})$ is a Wannier function product, $C^{(2)}_0(\mathbf{R})$ is an unnormalized spherical harmonic, and $P(\bar{a})$ is a regularized $s$-($p$-)wave pseudopotential~\cite{idziazek_Calarco_06} for bosons (fermions) which is assumed to have the universal dependence on the van der Waals length $\bar{a}$ discussed in Ref.~\cite{Idziaszek_Julienne_10,JMI}.  We evaluate the integrals Eq.~\eqref{eq:geo} using the ab initio method of Ref.~\cite{Wall_Carr_CE}.  As shown in that work, the dipole-dipole interaction parameters $U_{j-i,\sigma\sigma'}$ and $E_{j-i,\sigma_1\sigma_2\sigma_2'\sigma_1'}$ have a $\sim 1/|j-i|^p$ power-law tail with $p\gtrsim 3$ and a possibly significant exponentially decaying contribution when the lattice strengths are anisotropic.

\begin{table}
\caption{\label{table:fewbodyparams} The few-molecule parameters can differ by several orders of magnitude for experimentally relevant $^1\Sigma$ molecules~\cite{Aldegunde,Hfs,JMI}.  In order, we give the permanent dipole moment $d$, the nuclear spins $I_1$ and $I_2$, the rotational constant $B_N$, the nuclear quadrupole coupling constants $(eQq)_1$ and $(eQq)_2$, the nuclear $g$-factors $g_1$ and $g_2$, the hyperfine spin couplings $c_1$, $c_2$, $c_3$, and $c_4$, and the magneto-association field $B$.}
\begin{tabular}{|c|c|c|c|c|c|}
\hline &\LiCs&\NaK&\RbCs&\KRb&\LiNa\\
\hline $d$/Debye&5.52&2.76&1.25&0.566&0.56\\
\hline $I_1$&1&3/2&3/2&4&1\\
\hline $I_2$&7/2&4&7/2&3/2&3/2\\
\hline $B_N$/GHz&6.5202&2.8268&0.504&1.114&12.7355\\
\hline $\frac{(eQq)_1}{\mathrm{MHz}}$
&3$\times 10^{-4}$&-0.167&-0.872&0.45&8$\times 10^{-4}$\\[0.9ex]
\hline $\frac{(eQq)_2}{\mathrm{MHz}}$
&0.187&0.796&0.051&-1.41&0.687\\[0.9ex]
\hline $g_1$&0.822&1.479&1.834&-0.324&0.822\\
\hline $g_2$&0.738&-0.324&0.738&1.834&1.479\\
\hline $c_1$/Hz&15.2&110.6&98.4&-24.1&83.2\\
\hline $c_2$/Hz&3476.0&-90.9&194.1&420.1&802.2\\
\hline $c_3$/Hz&53.1&-46.5&192.4&-48.2&196.2\\
\hline $c_4$/Hz&620.8&-466.2&17345.4&-2030.4&212.3\\
\hline $B$/Gauss&843.5&139.7&181.7&545.9&745\\
\hline
\end{tabular}
\end{table}

\subsection{Tunability of MHH parameters for two internal states}
The single particle Hamiltonian consists of kinetic, lattice, and internal components, and the internal contribution has been discussed in detail in past work~\cite{Old_MHH,HMHH}.  We use the few-molecule parameters in Table~\ref{table:fewbodyparams} which have been calculated from density functional theory~\cite{Aldegunde} or available experimental data~\cite{Hfs}.  The internal states of an alkali dimer molecule in the presence of a DC electric field $\mathbf{E}_{\mathrm{DC}}=E_{\mathrm{DC}}\mathbf{e}_z$ group into manifolds $|\bar{N},M_N,\eta\rangle$. Here, $\bar{N}$ denotes that the manifold adiabatically correlates with having $N$ quanta of rotation as $E_{\mathrm{DC}}\to 0$, $M_N$ is the projection of rotation along $\mathbf{e}_z$, and $\eta$ indexes the nuclear hyperfine degeneracy.   Two such manifolds with $|M_N|\ne |M_N'|$ are well separated in energy provided that the scaled DC field strength $\beta_{\mathrm{DC}}\equiv dE_{\mathrm{DC}}/B_N\gtrsim 0.2$.  In addition to $\mathbf{E}_{\mathrm{DC}}$, we consider an applied DC magnetic field $\mathbf{B}=B\mathbf{e}_z$.  The values of $B$ we use, collected in Table~\ref{table:fewbodyparams}, correspond to the positions of the Feshbach resonances used for magneto-association~\cite{NaK,RbCs,KRb,LiCs,LiNa}; these are the magnetic fields naturally and easily used in experiments.  The Zeeman splittings and intrinsic couplings between hyperfine sublevels produce intra-manifold splittings which are typically a few tens to hundreds of kHz.

In order to induce rotational transitions between states in manifolds $|\bar{0},0,\eta\rangle$ and $|\bar{1},\pm1,\eta'\rangle$, we require a single AC microwave field with frequency $\omega$ and circular polarization with respect to the DC field axis; such fields are already used in experiments~\cite{Hfs}.  The detuning $\Delta_{\sigma}$ denotes how far from a single-molecule resonance state $\sigma$ is, and the effective Rabi frequency $\Omega_{\sigma\sigma'}$ expresses the strength of transitions from state $\sigma$ to state $\sigma'$.  In strong $B$ fields where the intra-manifold splitting is typically larger than e.g. the dipole-dipole interaction energy, the AC field determines which states are dynamically accessible.  Hence, the number of states involved in the MHH dynamics can be modified by the static field configuration and the frequency and power of the AC field, resulting in a tunable quantum complex system~\cite{Old_MHH}.  The number of dynamically accessible internal states can be quite sizable, typically a few tens, due to a large hyperfine degeneracy, see $I_1$ and $I_2$ in Table~\ref{table:fewbodyparams}.

\begin{figure}[tbp]
\centerline{\includegraphics[width=0.75\columnwidth]{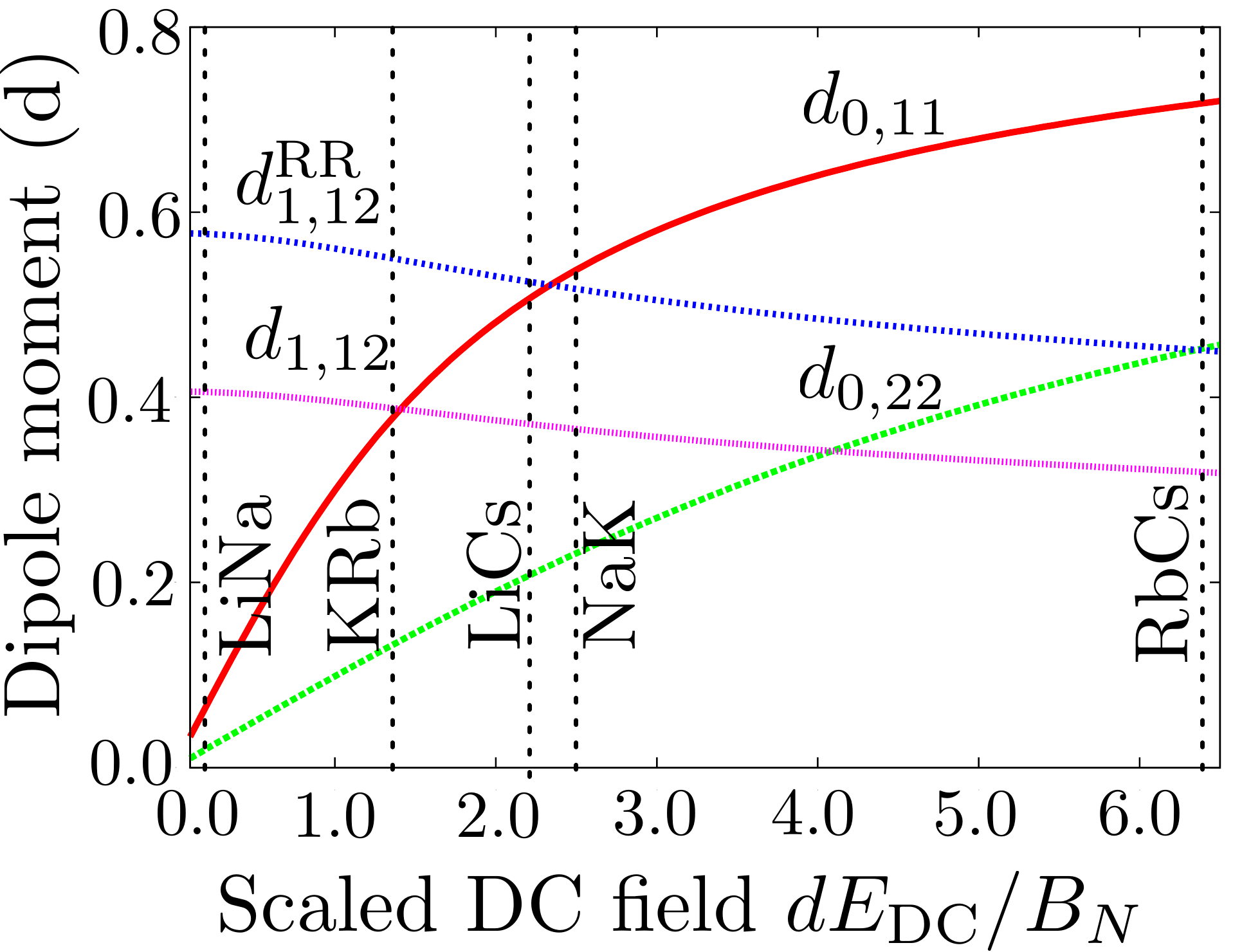}}
\caption{\label{fig:schematic}(Color online)  \emph{Examples of experimental controls not available to atoms.} An applied electric field $E_{\mathrm{DC}}$ gives rise to single-particle states with tunable expected dipole moments.  Vertical dashed lines indicate the position of $E_{\mathrm{DC}}=5$kV/cm for the various molecules. The expected transition dipole moments, giving rise to exchange interactions, are significantly different in the presence of an optical lattice ($d_{1,12}$, magenta short-dashed line) and in free space ($d_{1,12}^{\mathrm{RR}}$, dashed blue line).}
\vspace{-0.2in}
\end{figure}

Finally, we consider that the molecules are subject to a simple cubic optical lattice potential consisting of three pairs of counter-propagating laser beams, $V_{\mathrm{latt}}(\mathbf{r})=\textstyle\sum_{\nu\in\{x,y,z\}}E_{\eta}^2\cos^2(\pi \nu/a)\boldsymbol{\epsilon}_{\nu}^{\star}\cdot \tilde{\alpha} \cdot \boldsymbol{\epsilon}_{\nu}$.  Here, $\tilde{\alpha}$ is the polarizability tensor of the molecules~\cite{Kotochigova_DeMille_10}, $E_{\nu}^2$ is the intensity of the laser field along direction $\nu$, $\boldsymbol{\epsilon}_{\nu}$ is the polarization of the field along $\nu$, and $a=545$nm is the lattice spacing.  In the numerical examples we choose the polarizations $\boldsymbol{\epsilon}_x=\mathbf{e}_{y}$, $\boldsymbol{\epsilon}_y=\boldsymbol{\epsilon}_z=\mathbf{e}_{x}$ corresponding to current experiments~\cite{chotia_neyenhuis_11}.  As the tunneling depends on the particular rotational state, we define lattice heights as the lattice height corresponding to the absolute ground state, e.g., $V_x\equiv \langle \sigma=1| E_{x}^2\boldsymbol{\epsilon}_{x}^{\star}\cdot \tilde{\alpha} \cdot \boldsymbol{\epsilon}_{x}|\sigma=1\rangle$.  Here, the recoil energy $E_R=\hbar^2\pi^2/2ma^2$ and $|\sigma=1\rangle$ denotes the absolute ground state.  Due to the fact that the molecular polarizability tensor is rank two, light whose polarization is not along a pure spherical direction couples together the states in the manifold $|\bar{1},\pm 1,\eta\rangle$~\cite{Kotochigova_DeMille_10,Neyenhuis_12} and the tunneling amplitude $t_{p,\sigma}$ depends on the internal state and the polarization, see Table~\ref{table:MBparams}.  The relative tunneling amplitudes in different internal states can also be altered by changing the strength of the DC field or changing the frequency of the AC field to select driving to different internal states.  An additional consequence of the off-diagonal lattice coupling in the manifold $|\bar{1},\pm 1,\eta\rangle$ is that the single-particle eigenstates are in general a quasimomentum-dependent superposition of the states which diagonalize the internal Hamiltonian in free space, and so the dipole moments generally cannot be taken outside the integration in Eq.~\eqref{eq:geo}.

\begin{table}
\caption{\label{table:MBparams} Order of magnitude differences in Hubbard parameters give rise to a variety of phases.  Here we collect the many-body parameters of the MHH Eq.~\eqref{eq:MHH} for $V_y=V_z=55E_R$, $V_x=8E_R$, the few-molecule parameters given in Table~\ref{table:fewbodyparams}, and the $E_{\mathrm{DC}}$ and species given in the table.  The value $E_{\mathrm{DC}}=5$kV/cm is a typical strong field in experiments.  We do not include $\Omega$, as it is widely tunable by adjusting the power of the applied AC microwave field.}
\begin{tabular}{|c|c|c|c|c|c|}
\hline &\LiCs&\NaK&\RbCs&\KRb&\LiNa\\ 
\hline \multicolumn{6}{|c|}{$E_{\mathrm{DC}}$= 1 kV/cm}\\
\hline $U_{1,11}$ (kHz)&0.70&0.218&0.253&0.002&2$\times 10^{-5}$\\
\hline $U_{1,22}/U_{1,11}$&0.084&0.085&0.108&0.082&0.08\\
\hline$U_{1,12}/U_{1,11}$ &0.28&0.29&0.32&0.28&0.28\\
\hline$t_{1,1}/U_{1,11}$ &0.33&2.36&0.58&93.74&5$\times 10^{5}$\\
\hline$t_{1,2}/U_{1,22}$ &1.76&12.37&2.25&508.9&3$\times 10^{6}$\\
\hline $|E_{1,1212}/U_{1,11}|$&7.93&6.05&0.99&20.97&2$\times 10^3$\\
\hline \multicolumn{6}{|c|}{$E_{\mathrm{DC}}$= 5 kV/cm}\\
\hline $U_{1,11}$ (kHz)&8.97&2.48&0.970&0.049&5$\times 10^{-4}$\\
\hline $U_{1,22}/U_{1,11}$&0.14&0.16&0.35&0.11&0.08\\
\hline$U_{1,12}/U_{1,11}$ &0.38&0.40&0.59&0.32&0.28\\
\hline$t_{1,1}/U_{1,11}$ &0.026&0.20&0.15&5.13&2$\times 10^3$\\
\hline$t_{1,2}/U_{1,22}$ &0.068&0.48&0.14&19.75&1.2$\times 10^{5}$\\
\hline $|E_{1,1212}/U_{1,11}|$&0.519&0.43&0.18&1.05&114\\
\hline
\end{tabular}
\end{table}

The given experimental setup provides several controls which have no counterpart in atoms.  To provide explicit examples of this tunability, we now focus on a special case of the MHH in which only two internal states are populated, $\sigma\in\{1,2\}$, and the lattice is strongly confining in the $y$ and $z$ directions.  The condition of only two internal states is well-realized for current experiments with large $B$ fields and weak microwave power~\cite{KRb,Hfs}.  The state $|\sigma=1\rangle$ is taken to be the absolute ground state, and the state $|\sigma=2\rangle$ is the lowest energy eigenstate which maximizes the $\hat{d}_1$ transition dipole moment with $|\sigma=1\rangle$.  

Figure~\ref{fig:schematic} shows the dependence of the dipole moments $d_{q,\sigma\sigma'}$, see Eq.~\eqref{eq:Omega}, on the scaled DC field strength $\beta_{\mathrm{DC}}$.  The resonant moments $\{d_{0,\sigma\sigma}\}$, which are the expected dipole moments of the states $\{|\sigma\rangle \}$, give rise to the direct dipole-dipole interactions $U_{j-i,\sigma\sigma'}$, and the transition moments $\{d_{1,\sigma\sigma'}\}$,which give the strength of a dipole-allowed transition $\sigma\to\sigma'$, yield the exchange terms $E_{j-i,\sigma_1\sigma_2\sigma_2'\sigma_1'}$.  The transition moments are strongly affected by hyperfine structure and lattice polarization effects as shown by the discrepancy between $d_{1,12}$ and $d_{1,12}^{\mathrm{RR}}$, the latter being computed in the absence of a lattice.  

\section{Many-body case studies}
To demonstrate the emergent many-body features of Eq.~\eqref{eq:MHH} in near-term experimental setups, we present many-body case studies employing the infinite-size variational ground state search using matrix product states (iMPS) for 1D systems.  We choose the energetic cutoff defining $r_t$ etc.~in Eq.~\eqref{eq:MHH} to be 1Hz, and the field configurations are chosen such that terms not included in Eq.~\eqref{eq:MHH} such as assisted tunneling and tunneling outside of the quasi-1D domain fall below this threshold.  These cutoffs account for the fact that many-body phases induced by long-range interactions may be fragile against other terms in the Hamiltonian, non-conservative effects such as heating from the optical lattice, and lifetime constraints of the ultracold gas.

\subsection{Infinite-size variational ground state search using matrix product states}
Infinite-size variational ground state search using matrix product states (iMPS) is a numerical method which finds a representation for the ground state of an infinite one-dimensional lattice model~\cite{iMPS,Thesis}.  The state is assumed to be invariant under shifts by a user-specified number of lattice sites $q$, and hence can be represented in terms of a periodically repeating unit cell with $q$ lattice sites.  The unit cell is decomposed as a matrix product state (MPS) with finite entanglement cutoff $\chi$ measured by the Schmidt rank~\cite{SR}, and this unit cell is iteratively optimized via the solution of an eigenvalue problem.  The optimization reaches an end when the density matrix $\hat{\rho}$ obtained by tracing over all sites to the right of a given point in the infinite lattice changes by less than a tolerance $\epsilon$ between successive iterations.  We measure that change via the \emph{orthogonality fidelity}
\begin{align}
F_{\mathrm{ortho}}&\equiv \mathrm{Tr}\sqrt{\sqrt{\hat{\rho}_n}\hat{\rho}_{n-1}\sqrt{\hat{\rho}_n}}\, ,
\end{align}
introduced in Ref.~\cite{iMPS}.  Here, the subscript on $\hat{\rho}$ denotes the optimization cycle index.  In practice, we choose the orthogonality fidelity tolerance to be $\epsilon=\bar{\epsilon}_{\mathrm{local}}$, where $\bar{\epsilon}_{\mathrm{local}}$ is unit-cell averaged 2-norm distance between the optimal unit cell and its representation as an MPS with fixed entanglement cutoff $\chi$.  That is, deviations of the density matrix $\hat{\rho}$ between successive iterations are only from the errors due to a finite cutoff of entanglement.  Once the unit cell has converged, a standard orthonormalization procedure is used to define a normalized state which conforms to a known canonical MPS form on the infinite lattice~\cite{Schollwoeck_11}.  From this state, any desired observable may be computed.  Long-range interactions are accommodated by representing the Hamiltonian as a matrix product operator (MPO)~\cite{MPO}.  When the interaction cutoff $r_U$ is less than 6, we use a finite-range interaction on $r_U$ sites, but for larger cutoffs we use an infinite-range Hamiltonian obtained by fitting the function $U_{p}$ to a series of exponentials~\cite{NJP}.  We do this out of numerical convenience; the infinite-range decomposition requires fewer numerical resources for the same level of accuracy when $r_U>6$.

\subsection{Single-component phase diagram}
\begin{figure*}[htbp]
\centerline{\includegraphics[width=1.5\columnwidth]{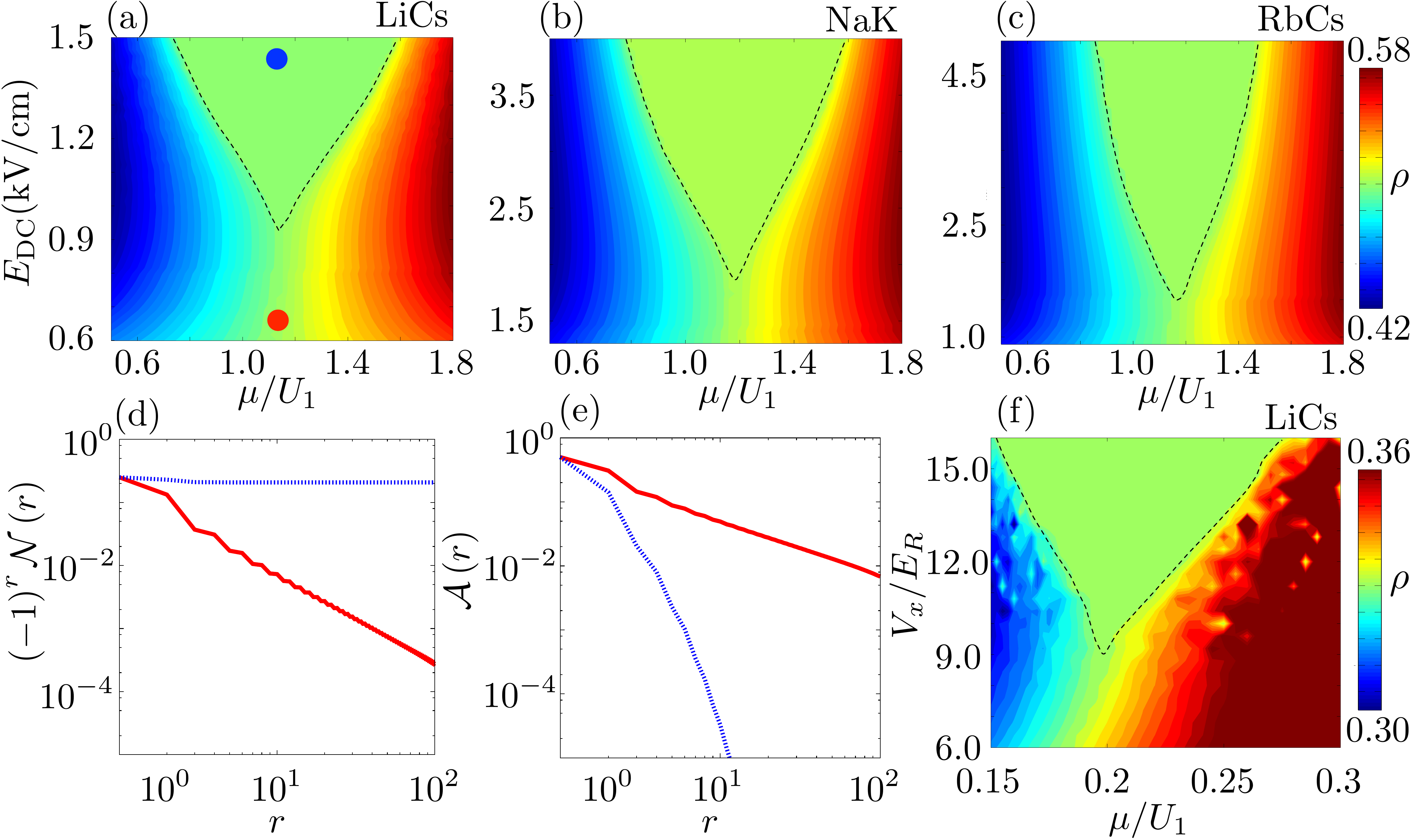}}
\caption{\label{fig:Omega0}(Color online)  \emph{Phase diagram of the MHH in the absence of an AC field.} (a) Density phase diagram for LiCs, showing the $\rho=1/2$ crystalline phase (CP) with non-vanishing single-particle gap.   The dashed line is a guide to the eye for the CP.  (b,c) analogs of (a) for NaK and RbCs, respectively.  (d) Density correlation function $\mathcal{N}(r)$ with period-two oscillation removed for the points marked in (a). (e) Single-particle density matrix $\mathcal{A}(r)$ for the points marked in (a).  (f) Density phase diagram for LiCs near the $\rho=1/3$ CP transition at fixed $E_{\mathrm{DC}}=3.5$kV/cm showing small density fluctuations induced by our use of the grand canonical ensemble.}
\vspace{-0.2in}
\end{figure*}

We first consider the phase diagram of Eq.~\eqref{eq:MHH} in a quasi-1D geometry and absent an AC field.  Without the AC field only the absolute ground state $|\sigma=1\rangle$ is populated, and so the MHH reduces to
\begin{align}
\label{eq:MHHACoff}  \hat{H}=\textstyle-\sum_{i\ne j;r_t}t_{j-i}\hat{a}_i^{\dagger}\hat{a}_{j} +\frac{1}{2}\sum_{i\ne j;r_U}U_{j-i}\hat{n}_{i}\hat{n}_{j}\, .
\end{align}
For bosonic {\RbCs}, the hard-core constraint is enforced by strong on-site dipole-dipole interactions $U_0\sim$4-20 kHz for the parameters of Table~\ref{table:MBparams}.  At half filling, $\rho=1/2$, a phase transition occurs at strong coupling (small $t_1/U_1$) to a crystalline phase (CP), see Fig.~\ref{fig:Omega0}.  Here, the CP corresponds to the region of fixed density which features with long-range order in the density correlation function $\mathcal{N}(r)\equiv \langle \Delta \hat{n}_0\Delta\hat{n}_r\rangle$, $\Delta\hat{n}_i=\hat{n}_i-\rho$; an exponential decay of the single-particle density matrix $\mathcal{A}(r)\equiv \langle \hat{a}_0^{\dagger}\hat{a}_r\rangle$; and a non-vanishing single-particle gap.  To within numerical resolution, we find that the CP melts directly into a superfluid with algebraic decay of both $\mathcal{N}(r)$ and $\mathcal{A}(r)$ and no single-particle gap, and there is no region of supersolid phase displaying simultaneous long-range order in $\mathcal{N}(r)$ and algebraic decay of $\mathcal{A}(r)$.  Correlation functions for the red (lower) and blue (upper) points in Fig.~\ref{fig:Omega0}(a) are shown as the solid red and dashed blue curves in Figs.~\ref{fig:Omega0}(d) and (e).  For situations where the hard-core condition is relaxed, other phases induced by assisted and pair tunneling may appear~\cite{Sowinski}.  Due to their small dipole moment, {\KRb} and {\LiNa} do not display the CP phase transition for fields $E_{\mathrm{DC}}\lesssim 5$kV/cm, and so are not included in Fig.~\ref{fig:Omega0}.  The absence of the CP transition for these molecules is an example of how different molecules display different many-body physics.

\begin{figure}[h]
\centerline{\includegraphics[width=0.7\columnwidth]{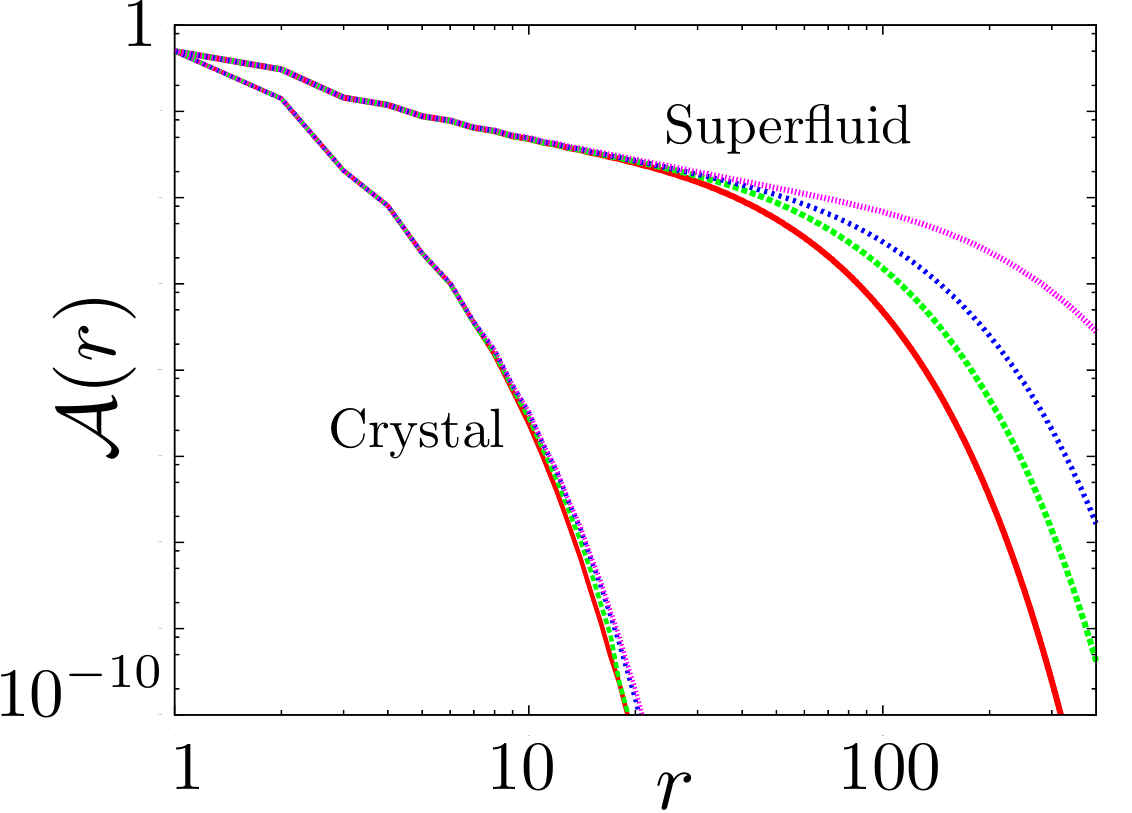}}
\caption{\label{fig:suppl}  (Color online) \emph{Scaling of single-particle density matrix with entanglement cutoff in iMPS.}  The single-particle density matrix $\mathcal{A}(r)$ computed by iMPS for $\chi=$12 (red solid line), 16 (green dashed line), 20 (blue dotted line), and 40 (purple fine-dashed line) in the crystalline and superfluid phases.  In the crystalline phase, $\mathcal{A}(r)$ decays exponentially and so the results converge for finite $\chi$.  In the superfluid phase $\mathcal{A}(r)$ has an algebraic decay, and the range over which this decay is well-approximated increases with $\chi$.}
\end{figure}

After optimizations have been performed with a range of entanglement cutoffs $\chi$, the effects of finite entanglement cutoff can be discerned by a scaling analysis in $\chi$.  It is known rigorously that MPSs with a finite cutoff $\chi$ cannot display power-law decay of correlation functions for arbitrary distances.  Hence, for a given $\chi$, there exists a distance $r_{\chi}$ beyond which a correlation function with a power-law decay begins to decay exponentially, and $r_{\chi}$ is an increasing function of $\chi$.  In addition, the scaling for some universal quantities, such as the correlation length, have a known scaling relation with $\chi$ which depends only on the data of the conformal field theory describing criticality in the infrared limit~\cite{Scaling}.  An example of the scaling of correlation functions with $\chi$ is given in Fig.~\ref{fig:suppl}, which shows the scaling of the single-particle density matrix $\mathcal{A}(r)$ in the superfluid and crystalline phase near half filling.  The results in Fig.~\ref{fig:Omega0} have been checked for convergence in $\chi$ and $q$, the unit cell length.

When the interaction $U_{x}$ is a convex function of $x$, regions of rational density other than $\rho=1/2$ occupy finite regions in the phase diagram~\cite{Luttingerstaircase}.  To study the accessibility of such phases, we investigate the CP at filling $\rho=1/3$ for {\LiCs} in Fig.~\ref{fig:Omega0}(f).  Here we fix $E_{\mathrm{DC}}$ and vary the quasi-1D lattice height $V_x$ to change $t_1/U_1$.  While it is possible to achieve this phase with the other molecules, the equilibration timescales will be longer than the lifetime of typical experiments due to the smaller dipole interaction energy which sets the timescale, see Table~\ref{table:MBparams}.  Hence, {\LiCs} is the best choice for investigating phases due to strong direct interactions.  Similarly, other CPs with densities $\rho<1/3$ could be seen in principle, but the necessity of very small $t_1/U_1$ requires equilibration times and temperatures which are challenging to achieve.  We note that the $\rho=1/3$ CP is unique to molecules with long-range dipole-dipole interactions; there is no analog of this phase in atoms with short-range interactions.  

The density plot near the $\rho=1/3$ crystalline phase, Fig.~\ref{fig:Omega0}(f), appears noisy.  This apparent noise is an artifact of having a finite resolution of points in the plot and using the grand canonical ensemble.  The density fluctuation spots become smaller for higher resolution, and are concentrated on the border between two regions of fixed density.  Hence, as the resolution increases, the effect of these fluctuations will become less prominent.  Density fluctuations on the border between superfluid regions of fixed density are also seen for finite systems in the grand canonical ensemble, but do not appear in the canonical ensemble~\cite{Meso}.  Finite-size effects introduce a non-vanishing single-particle gap, and this inherently limits the resolution of the chemical potential for the grand canonical ensemble in a finite-volume system.


\subsection{Multi-component physics}
The simplest situation utilizing the internal structure of the molecule is in a deep lattice with one molecule per lattice site where tunneling is negligible and only two internal states are accessible.  When the two internal states of the molecule are mapped to an effective two-component spin via the standard transformation $\hat{S}^z_i\equiv (\hat{n}_{1 i}-\hat{n}_{2i})/2$, $\hat{S}^{+}_i\equiv \hat{a}_{1i}^{\dagger}\hat{a}_{2i}$, $\hat{S}^{-}_i=(\hat{S}^+_i)^{\dagger}$, the MHH for either bosons or fermions maps, up to constant terms, onto a long-range XXZ model in transverse and longitudinal fields
\begin{align}
\label{eq:XXZ} \hat{H}=&\textstyle \frac{1}{2}\sum_{i\ne j,r_X}X_{j-i}\hat{S}^+_i\hat{S}^-_j-2\Omega \sum_i \hat{S}^x_i+h_z\sum_i\hat{S}_i^z\\
\nonumber &+\textstyle\frac{1}{2}\sum_{i\ne j;r_I} I_{j-i}\hat{S}^z_i\hat{S}^z_j\, ,
\end{align}
{as has been derived by Gorshkov \emph{et. al.}~\cite{Gorshkov_Manmana} and studied in 1D without the $\Omega$ term~\cite{XXZ}.}  Here, $I_{j-i}\equiv U_{j-i,11}-2U_{j-i,12}+U_{j-i,22}$, $X_{j-i}\equiv E_{j-i,1212}$, and $h_z\equiv (\Delta_1-\Delta_2)+\sum_{j-i;r_U} (U_{j-i,11}-U_{j-i,22})$.  We stress that the ratios of the various spin couplings depend in an essential way on the molecular species, see Table~\ref{table:MBparams}.  {With further fine tuning, a wider range of the XXZ model may be accessed~\cite{Gorshkov_Manmana}.}

\section{Conclusions}
We have presented the molecular Hubbard Hamiltonian, the natural Hamiltonian governing the many-body physics of ultracold molecules in optical lattices.  Molecules are sensitive to experimental controls which are not relevant for atoms such as an applied DC electric field, the polarization of optical lattice light, and the frequency and power of an AC microwave field driving rotational transitions.  The response of a molecule to external fields is also much more dependent on molecular species than is the case for atoms.  Our results establish that the molecules employed in current experiments will display a range of species-dependent many-body features.

We acknowledge useful discussions with J.~Aldegunde, J.~L.~Bohn, A.~V.~Gorshkov, J.~M.~Hutson, K.~Maeda, and M.~Weidem\"uller.  This work was supported by the Alexander von Humboldt Foundation, the Heidelberg Center for Quantum Dynamics, AFOSR grant number FA9550-11-1-0224, and the National Science Foundation under Grants PHY-1207881, PHY-1067973,  PHY-0903457, and PHY11-25915.  We also acknowledge the Golden Energy Computing Organization at the Colorado School of Mines for the use of resources acquired with financial assistance from the National Science Foundation and the National Renewable Energy Laboratories.

\bibliographystyle{prsty}

\end{document}